\documentclass[runningheads]{llncs}
\usepackage{graphicx}
\usepackage{color}
\usepackage{xcolor,colortbl}
\usepackage{pifont}
\usepackage{adjustbox}
\usepackage{array}
\usepackage{cite}
\usepackage{multirow}
\usepackage{booktabs}
\usepackage{mathtools}
\usepackage{amssymb}
\usepackage{amsmath}
\usepackage{mathabx} 
\usepackage{tikz}

\newcommand{\ra}[1]{\renewcommand{\arraystretch}{#1}}
\newcolumntype{C}[1]{>{\centering\let\newline\\\arraybackslash\vspace{0pt}}m{#1}}
\usepackage{array, multirow, bigdelim, makecell, booktabs} 
\usetikzlibrary{arrows,positioning,shapes.geometric}
\definecolor{lightblue}{RGB}{153,204,255}
\definecolor{lightgreen}{RGB}{53,255,153}
\usepackage{lipsum}
\usetikzlibrary{arrows,decorations.markings}
\usetikzlibrary{arrows.meta}
\usepackage{figsize}
\usepackage[T1]{fontenc}
\usepackage{subfigure}

\begin{document}
\title{Building Brains: Subvolume Recombination for Data Augmentation in Large Vessel Occlusion Detection} %
\author{Florian Thamm \inst{1,2} \and Oliver Taubmann \inst{2} \and Markus Jürgens \inst{2} \and Aleksandra Thamm \inst{1} \and Felix Denzinger \inst{1,2} \and Leonhard Rist \inst{1,2} \and  Hendrik Ditt \inst{1}\and Andreas Maier \inst{1}}

\titlerunning{Subvolume Recombination for Augmentation in LVO Detection}
\authorrunning{F. Thamm et al. \inst{1}\orcidID{0000-1111-2222-3333} }
\institute{Friedrich-Alexander University Erlangen-Nuremberg, Erlangen, Germany \and Siemens Healthcare GmbH, Forchheim, Germany}
\maketitle              %
\begin{abstract}
Ischemic strokes are often caused by large vessel occlusions (LVOs), which can be visualized and diagnosed with Computed Tomography Angiography scans. As time is brain, a fast, accurate and automated diagnosis of these scans is desirable.
Human readers compare the left and right hemispheres in their assessment of strokes. A large training data set is required for a standard deep learning-based model to learn this strategy from data. As labeled medical data in this field is rare, other approaches need to be developed.
 To both include the prior knowledge of side comparison and increase the amount of training data, we propose an augmentation method that generates artificial training samples by recombining vessel tree segmentations of the hemispheres or hemisphere subregions from different patients. The subregions cover vessels commonly affected by LVOs, namely the internal carotid artery (ICA) and middle cerebral artery (MCA). In line with the augmentation scheme, we use a 3D-DenseNet fed with task-specific input, fostering a side-by-side comparison between the hemispheres. Furthermore, we propose an extension of that architecture to process the individual hemisphere subregions. All configurations predict the presence of an LVO, its side, and the affected subregion.
We show the effect of recombination as an augmentation strategy in a 5-fold cross validated ablation study. We enhanced the AUC for patient-wise classification regarding the presence of an LVO of all investigated architectures. For one variant, the proposed method improved the AUC from 0.73 without augmentation to 0.89. The best configuration detects LVOs with an AUC of 0.91, LVOs in the ICA with an AUC of 0.96, and in the MCA with 0.91 while accurately predicting the affected side.

\keywords{Computed Tomography Angiography  \and Stroke \and Augmentation}
\end{abstract}

\section{Introduction}
A multitude of branches and bifurcations characterize the vascular structure in the brain. Once this system is disturbed, focal deficits may be the consequence. Ischemic stroke is the most common cerebrovascular accident, where an artery is occluded by a clot preventing blood flow. If in particular large vessels are affected, this condition is called large vessel occlusion (LVO), which primarily appears in the middle cerebral artery (MCA) and/or internal carotid artery (ICA). Computed Tomography Angiography (CTA) is commonly used for diagnosis, for which contrast agent is injected to enhance all perfused vascular structures. LVOs in the anterior circulation become visible in such images as unilateral discontinuation of a vessel. Therefore, human readers take symmetries into account and usually compare the left and right hemispheres to detect LVOs. Despite common anatomical patterns, the individual configuration and appearance of the vessel tree can differ substantially between patients, hence automated and accurate methods for LVO detection are desirable.

Amukotuwa et al.~developed a pipeline consisting of 14 image processing steps to extract hand-crafted features followed by a rule-based classifier. With their commercially available product, they achieved an area under the receiver operator characteristic curve (AUC) between 0.86 and 0.94 depending on the patient cohort \cite{3214-amukotuwa2019automated,3214-amukotuwa2019fast}. Multi-phase CTA which consists of scans acquired in the arterial, peak venous, and late venous phase has been used as well, as it offers more information about the underlying hemodynamics. %
Stib et al.~\cite{3214-stib2020detecting} computed maximum-intensity projections of the vessel tree segmentation of each phase and predicted the existence of LVOs with a 2D-DenseNet architecture~\cite{3214-huang2017densely}. They performed an ablation study on 424 patients using the individual phases and achieved AUCs between 0.74 and 0.85. Another commercially available detection tool for LVOs utilizing a Convolutional Neural Network (CNN) has been evaluated by Luijten et al.~\cite{3214-luijten2021diagnostic}, who report an AUC of 0.75 on a cohort of 646 patients.

In all studies mentioned above, large data pools have been used for development and evaluation. In contrast, Thamm et al. \cite{3214-thammlinger-bvm} proposed a method to counteract the problem of limited data while achieving a comparable performance as related work. They showed that the application of strong elastic deformations on vessel tree segmentation leads to significant improvements in terms of the detection of LVOs.
However, deformations only change vessel traces and do not vary the underlying tree topologies. Furthermore, vessels, especially the ICA, have asymmetric diameters in the majority of patients \cite{caplan2007arterial}, which does not get significantly altered by deformations. Finally, the image regions relevant for strokes are spatially distant from each other. Previous work did not consider this for the development of their architectures.

\begin{figure}[tb]
    \centering
    \includegraphics[width=\textwidth]{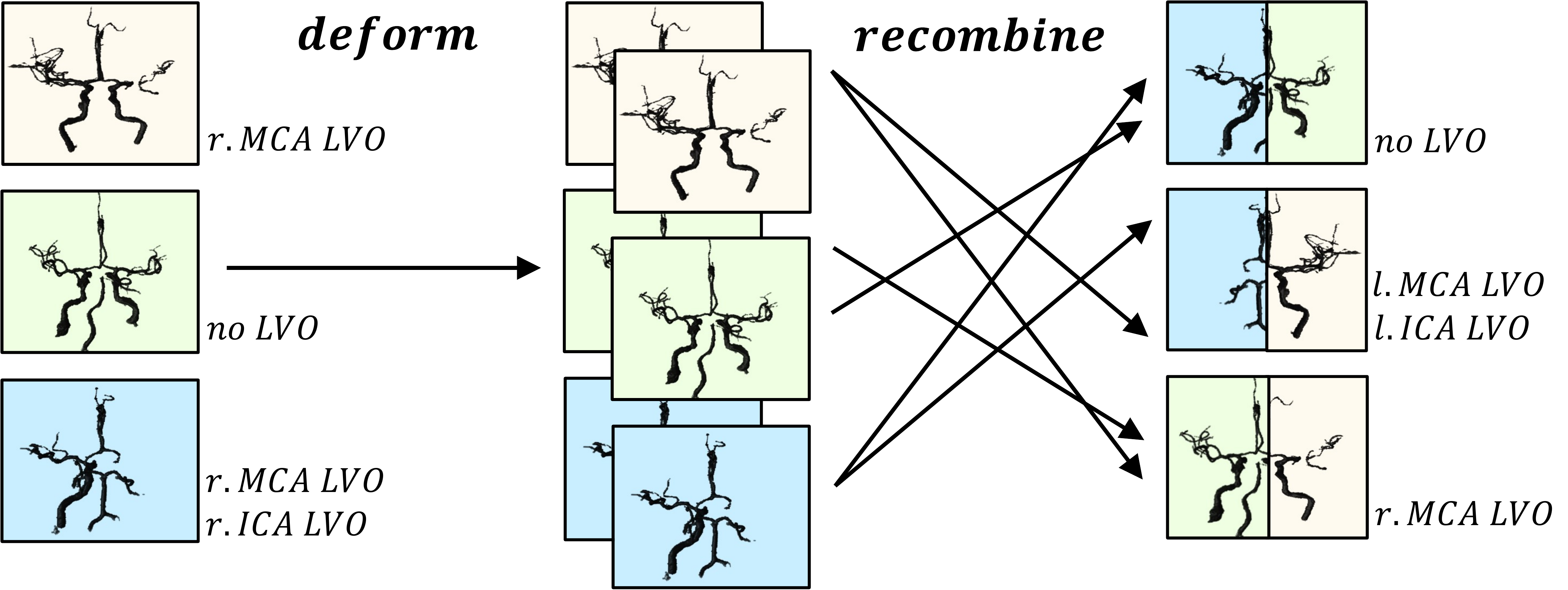}
    \caption{Proposed augmentation method. Vessel tree segmentations are deformed and randomly recombined yielding artificial patients.} 
    \label{img:recomb}
\end{figure}

Advancing the idea of data enrichment with information-preserving methods, we propose a randomized recombination of different patients as a novel augmentation method to create new synthetic data. This data is leveraged to train a 3D-DenseNet with a task-specific input feeding, which allows the network to perceive the vessel trees of both hemispheres simultaneously. This approach is extended by using two 3D-DenseNet encoding individual subregions (ICA, MCA) to further integrate prior knowledge. As the proposed augmentation neither requires additional memory nor delays training, it can supplemented with other more computational expensive methods like the elastic deformations which in related works have shown positive impact (Fig.~\ref{img:recomb}). Furthermore, we additionally predict if the left or right ICA and/or MCA vessels are occluded, which implicitly enables a coarse localization of the LVO.

\section{Methods}
\subsection{Data}
\label{sec:data}
The data used in this study consists of CTA scans covering the head/neck region of 151 patients, of which 44/57 suffered from left-/right-sided LVOs. The ICA was affected in 12/24 and the MCA in 40/50 cases for the left/right side, i.e.~in some cases ICA and MCA are affected. MCA LVOs are either located proximal/distal in M1 or in the transition from M1 to M2. The data was provided by one clinical site and acquired with a Somatom Definition AS+ (Siemens Healthineers, Forchheim, Germany).

\subsection{Preprocessing}
\label{sec:preprocessing}
For preprocessing, the cerebral vasculature is segmented following the approach of Thamm et al.~\cite{3214-thammlinger-vcbm}. Then the scans are non-rigidly registered using Chefd'hotel et al.'s method \cite{3214-chefd2002flows} into a reference coordinate system of a probabilistic brain atlas \cite{3214-kemmling2012decomposing} with isotropic voxel spacings of 1\,mm. The vessel segmentation is transformed into the reference coordinate system using the determined deformation field. As a result, the vessel trees have a uniform orientation and sampling. The reference coordinate system enables to split the volume easily into both hemispheres or to extract specific subregions of the segmentation.   
\newpage
\subsection{Hemisphere Recombination}
\label{sec:recombination-h}
To augment our data we leverage the fact that the human brain and therefore the cerebral vasculature is quasisymmetric w.r.t.~to the mid-plane dividing the brain into the left and right hemisphere. We therefore split the vessel tree segmentations of $P$ patients along the brain mid-plane. Next, we mirror all left-sided hemisphere's vessel trees in sagittal direction such that they become congruent to their right counterparts. A ``hemisphere tree'' describes in this context the segmented tree-like vasculature corresponding to one hemisphere. The set of hemisphere tree segmentations for all patients is denoted as $\mathcal{X}^{\text{H}} = \{\boldsymbol{x}_1^{\text{H}}, \dots, \boldsymbol{x}_N^{\text{H}}\}$ where $N = 2P$. Each hemisphere tree $\boldsymbol{x}^{\text{H}}$ of volume size $100 \times 205 \times 90$ voxels (mm$^3$) (atlas space split in half) is labeled regarding the presence of an LVO with $y^{\text{H}} \in \{0,1\}$ (1 if LVO positive, 0 else). Analogously, the set of all labels is denoted as $\mathcal{Y}^{\text{H}} = \{y_1^{\text{H}}, \dots, y_N^{\text{H}}\}$. 

New artificial patients can be generated by randomly drawing two hemisphere trees $\boldsymbol{x}_i^{\text{H}}$ and $\boldsymbol{x}_j^{\text{H}}$ from $\mathcal{X}^{\text{H}}$ which are recombined to a new stack $s_{ij} = \{\boldsymbol{x}_i^{\text{H}}, \boldsymbol{x}_j^{\text{H}}\}$. Recombinations of two LVO positive hemisphere trees lead to no meaningful representation and are hence excluded. If $r$ is the ratio of LVO positives cases among the $P$ patients, the augmentation scheme leads to 
\begin{equation}
    R = {\underbrace{\textstyle {2r(2-r)P^2}}_{\mathclap{\text{LVO pos.+Mirr}}}} + {\underbrace{\textstyle {(2-r)^2P^2}}_{\mathclap{\text{LVO neg.$+$Mirr}}}}
\end{equation}
possible recombinations including the mirrored representations. In this work, approximately 81k recombinations can be achieved using P=151 patient with r=67\% being LVO positive.

We consider a multi-class classification problem with one-hot encoded labels $\boldsymbol{y}'^{\text{H}} \in \{0,1\}^3$. The encoding comprises three exclusive classes using the following order: No LVO, left LVO or right LVO. For two recombined hemisphere trees, $i$ for left and $j$ for the right, the labels $y_i^\text{H}$, $y_j^\text{H}$ relate to $\boldsymbol{y}'^{\text{H}}_{ij}$ as 
\begin{equation}
    \boldsymbol{y}'^{\text{H}}_{ij} = \begin{bmatrix}
                             \neg ( y_i^{\text{H}} \lor  y_j^{\text{H}} ) \\
                              y_i^{\text{H}} \\
                             y_j^{\text{H}}
                          \end{bmatrix}.
\end{equation}
By stacking randomly selected hemisphere trees creating new artificial data sets, a neural network $f$ can be trained for all $i$ and $j$ such that $f(s_{ij}^{\text{H}}) = \hat{\boldsymbol{y}}'^{\text{H}}_{ij} \approx \boldsymbol{y}'^{\text{H}}_{ij} \quad \forall i,j \in \{1, \dots N\}$. The stacking can be done sagittally or channel-wise depending on the architecture (Sec.~\ref{sec:architectures}).

The recombination does neither have to be random and unstructured nor cover all permutations. Instead, the construction can be targeted to create specific distributions. In our case, we sample the three classes to be uniformly distributed, whereby each hemisphere appears once in one epoch.

\subsection{Recombination of ICA and MCA Subvolumes}
\label{sec:recombination-im}

The above-mentioned recombination strategy can be extended if the each hemisphere vessel segmentation is further partitioned in two subvolumes, respectively covering the ICA and MCA regions initially defined based on the atlas. Likewise with $\mathcal{X}^{\text{H}}$, the subvolumes of the hemisphere trees of all patients are described as the ICA set $\mathcal{X}^{\text{I}} = \{\boldsymbol{x}_1^{\text{I}}, \dots, \boldsymbol{x}_N^{\text{I}}\}$ with a size of $55 \times 121 \times 57$ voxels (mm$^3$) for $\boldsymbol{x}^{\text{I}}$, and the MCA set $\mathcal{X}^{\text{M}} = \{\boldsymbol{x}_1^{\text{M}}, \dots, \boldsymbol{x}_N^{\text{M}}\}$ with $x^{\text{M}}$ of size $60 \times 77 \times 76$ voxels (mm$^3$).
Each subvolume is associated with a corresponding label $\mathcal{Y}^{\text{I}}$ and $\mathcal{Y}^{\text{M}}$ analogously to $\mathcal{Y}^{\text{H}}$ described in Sec.~\ref{sec:recombination-h}.

For the synthesis of new artificial patients, subvolumes $\boldsymbol{x}_i^{\text{I}}$ and $\boldsymbol{x}_j^{\text{I}}$ are drawn from $\mathcal{X}^{\text{I}}$ and $\boldsymbol{x}_k^{\text{M}}$ and $\boldsymbol{x}_l^{\text{M}}$ from $\mathcal{X}^{\text{M}}$, forming a new stack $s_{ijkl} = \{\boldsymbol{x}_i^{\text{I}}, \boldsymbol{x}_j^{\text{I}}, \boldsymbol{x}_k^{\text{M}}, \boldsymbol{x}_l^{\text{M}}\}$ whereby asymmetric global LVOs (e.g.\,l.\,ICA and r.\,MCA) are excluded. A patient's one-hot encoded labels are not only represented globally by $\boldsymbol{y}'^{\text{H}}$ anymore but may be decomposed into $\boldsymbol{y}'^{\text{M}} \in \{0,1\}^3$ and  $\boldsymbol{y}'^{\text{I}} \in \{0,1\}^3$ identically representing the exclusive classes described above, restricted to their respective region:

\begin{figure}[tb]
  \centering
  \subfigure[ICA/MCA subvolumes in an angled perspective. \label{fig:box}]{\includegraphics[width=.47\textwidth]{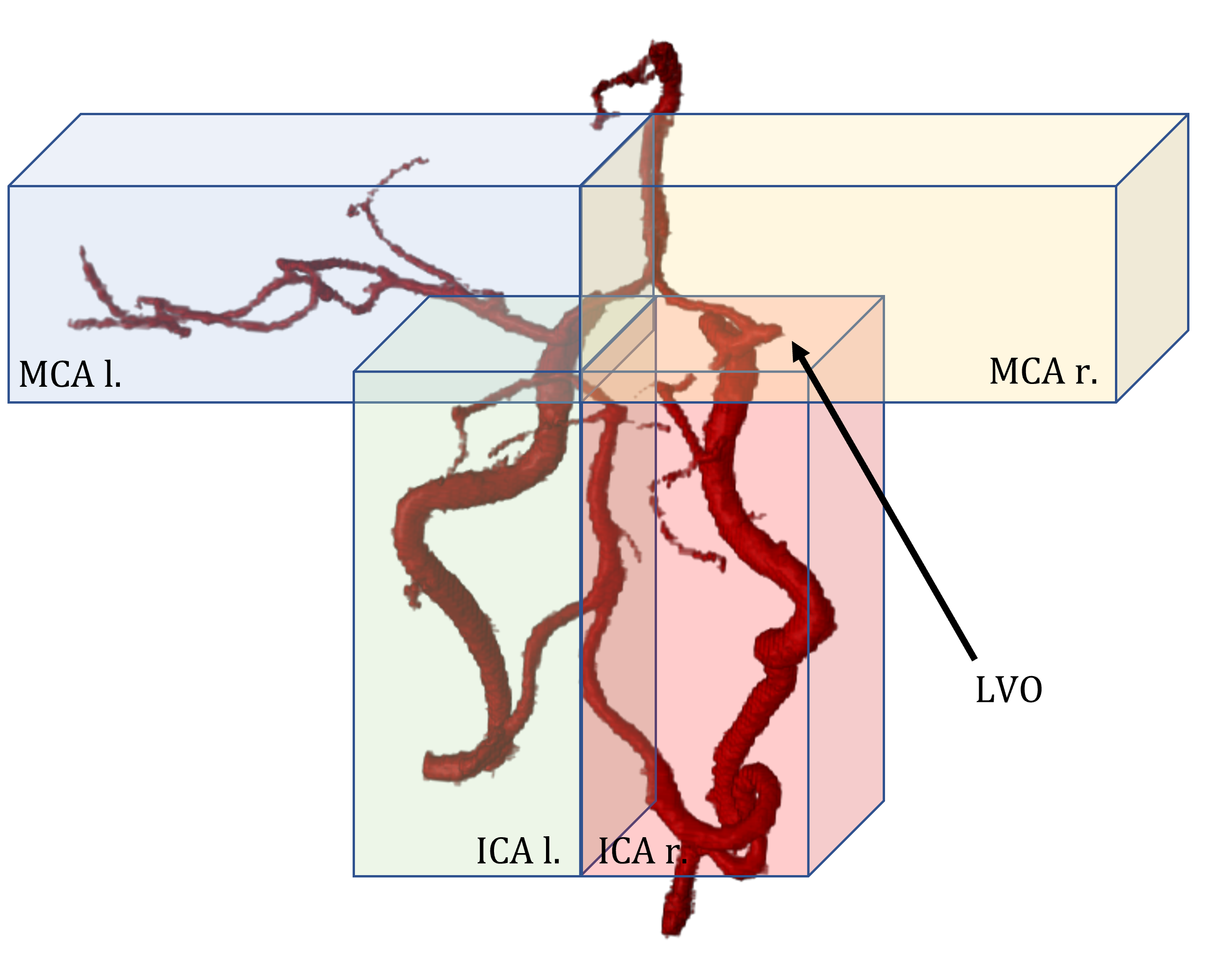}} \quad
  \subfigure[ICA/MCA subvolumes viewed axially caudal.\label{fig:box2}]{\includegraphics[width=.47\textwidth]{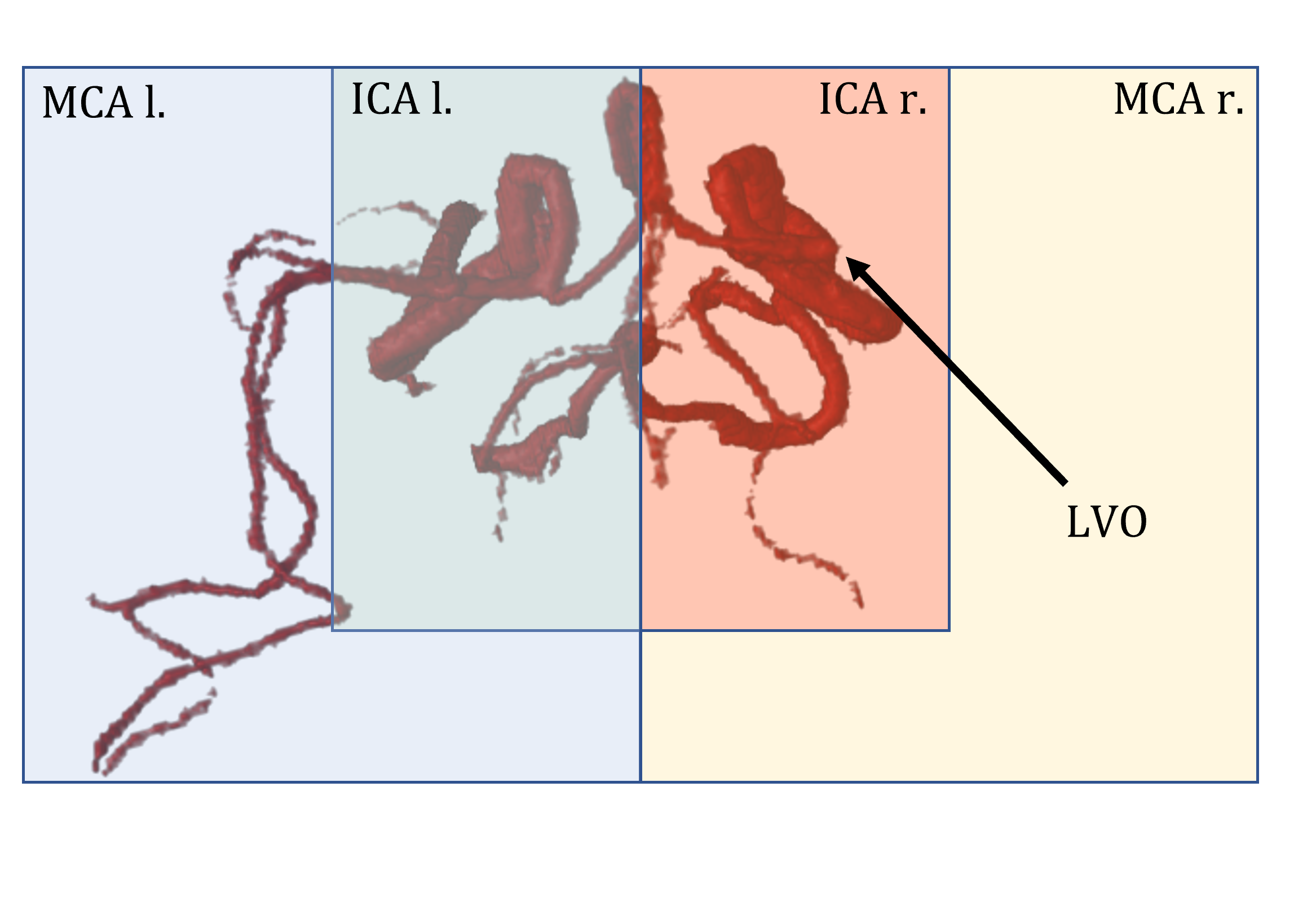}}
  \caption{Circle of Willis and surrounding vessels including the subvolumes covering the ICAs/MCAs each left and right with on colored overlay. This example shows an LVO on the right proximal MCA visible as an interruption in the respective vessel trace.}\label{mainfigure}
\end{figure} 
\begin{equation}
\label{eq:im_labels}
    \boldsymbol{y}'^{\text{H}}_{ijkl} = \begin{bmatrix}
                             \neg ( y_i^{\text{I}} \lor y_j^{\text{I}} \lor y_k^{\text{M}} \lor  y_l^{\text{M}} ) \\
                              y_i^{\text{I}} \lor y_k^{\text{M}}  \\
                              y_j^{\text{I}} \lor y_l^{\text{M}}
                          \end{bmatrix},
    \boldsymbol{y}'^{\text{I}}_{ij} = \begin{bmatrix}
                             \neg ( y_i^{\text{I}} \lor  y_j^{\text{I}} ) \\
                              y_i^{\text{I}} \\
                             y_j^{\text{I}}
                          \end{bmatrix}  ,                    
   \boldsymbol{y}'^{\text{M}}_{kl} = \begin{bmatrix}
                             \neg ( y_k^{\text{M}} \lor  y_l^{\text{M}} ) \\
                              y_k^{\text{M}} \\
                             y_l^{\text{M}}
                          \end{bmatrix}                  
\end{equation}
The suggested distinction between ICA and MCA within one hemisphere increases the number of possible recombinations. Assuming in all LVO positives, the ICA and MCA are affected, we obtain 
\begin{equation}
R = {\underbrace{\textstyle {2r^2(2-r)^2P^4+4r(2-r)^3P^4}}_{\mathclap{\text{LVO pos.+Mirr}}}} + {\underbrace{\textstyle {(2 - r)^4P^4}}_{\mathclap{\text{LVO neg.$+$Mirr}}}}
\end{equation}
recombinations. Analogously to Sec.~\ref{sec:architectures}, $~5730$M recombinations are possible in this work.

\subsection{Architectures}
\label{sec:architectures}
We make use of the 3D-DenseNet architecture \cite{3214-huang2017densely} such that it takes advantage of the proposed data representation by design. The first variant is trained on $\mathcal{X}^{\text{H}}$ and its extension is trained using both, $\mathcal{X}^{\text{I}}$ and $\mathcal{X}^{\text{M}}$. As baseline serves the approach by Thamm et al.~\cite{3214-thammlinger-bvm} utilizing 3D-DenseNets trained on whole heads which were not split into two hemispheres. However, all architectures predict $\hat{\boldsymbol{y}}'^{\text{M}}$, $\hat{\boldsymbol{y}}'^{\text{I}}$ and $\hat{\boldsymbol{y}}'^{\text{H}}$ (Eq.~\ref{eq:im_labels}).

\subsubsection{Baseline (Whole Head)}
We compare the suggested architectures and augmentation schemes to the methods presented in Thamm et al.~\cite{3214-thammlinger-bvm}. Each data set is deformed 10 times with a random elastic field using the parameters described in \cite{3214-thammlinger-bvm} and mirrored sagittally (left/right flip). A 3D-DenseNet \cite{3214-huang2017densely} ($\approx$\,4.6m parameters, growth rate of 32 and 32 initial feature maps) receives the vessel tree segmentations covering the entire head generated and preprocessed in the way described in Sec.~\ref{sec:preprocessing}. The recombination method is applicable here, if the two drawn hemispheres trees $\boldsymbol{x}_i^{\text{H}}$ and $\boldsymbol{x}_j^{\text{H}}$ are sagittally concatenated along the brain midplane. The indices for the labels are hence $\boldsymbol{y}'^{\text{H}}_{ijij}$, $\boldsymbol{y}'^{\text{I}}_{ij}$ and $\boldsymbol{y}'^{\text{I}}_{ij}$, using the notation described in Sec.~\ref{sec:recombination-im}. 

\subsubsection{Hemisphere-Stack (H-Stack)}
The first proposed variant is based on the 3D-DenseNet architecture ($\approx$\,3m parameters - larger capacity did not lead to better performances) as well and receives the samples as described in Sec.~\ref{sec:recombination-h} drawn from $\mathcal{X}^{\text{H}}$, but concatenated channel-wise (Fig.~\ref{fig:h-stack}). This has the advantage that the corresponding positions on both sides are spatially aligned within the same receptive fields of convolutional layers even early in the network and are carried over to deeper stages of the network due to the skip connections of the DenseNet architecture. Hence, a left/right comparison between both hemispheres is encouraged by design and does not need to be encoded over long spatial distances. The only difference between H-Stack and the baseline exists therein the feeding and composition of the data to demonstrate the impact of the suggested data representation. The label computation is identical. Furthermore, as mentioned in Sec.~\ref{sec:recombination-h} we extend the recombination with the elastic deformations using RandomElasticDeformation \cite{3214-torchio} (TorchIO, max.~displacement 20 voxels, 6 anchors, 10 repetitions).

\subsubsection{ICA-MCA-Stack (IM-Stack)}
\begin{figure}[tb]
  \centering
  \subfigure[H-Stack\label{fig:h-stack} utilizing a DenseNet to predict all 9 classes]{\includegraphics[width=.47\textwidth]{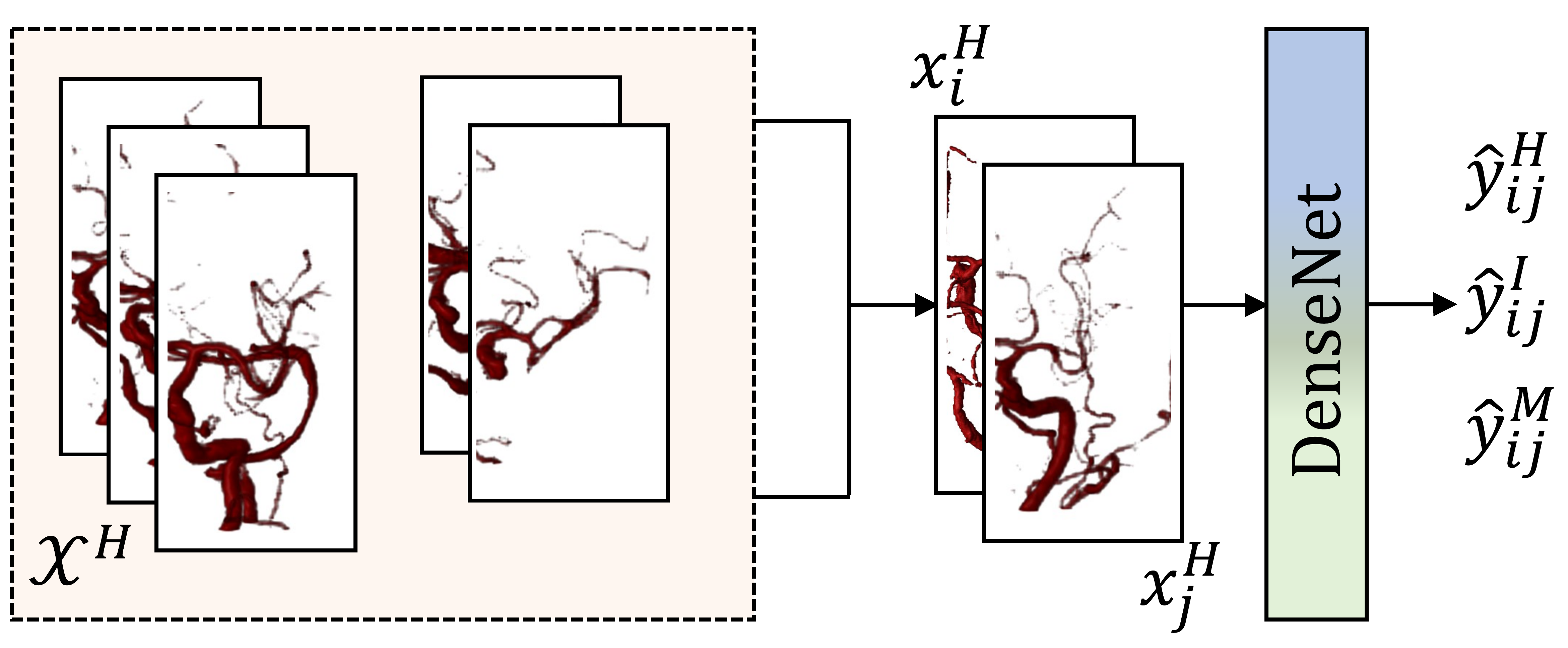}} \quad
  \subfigure[IM-Stack\label{fig:im-stack} consisting of two enconders each returning feature vectors which are processed by 3 classifiers]{\includegraphics[width=.47\textwidth]{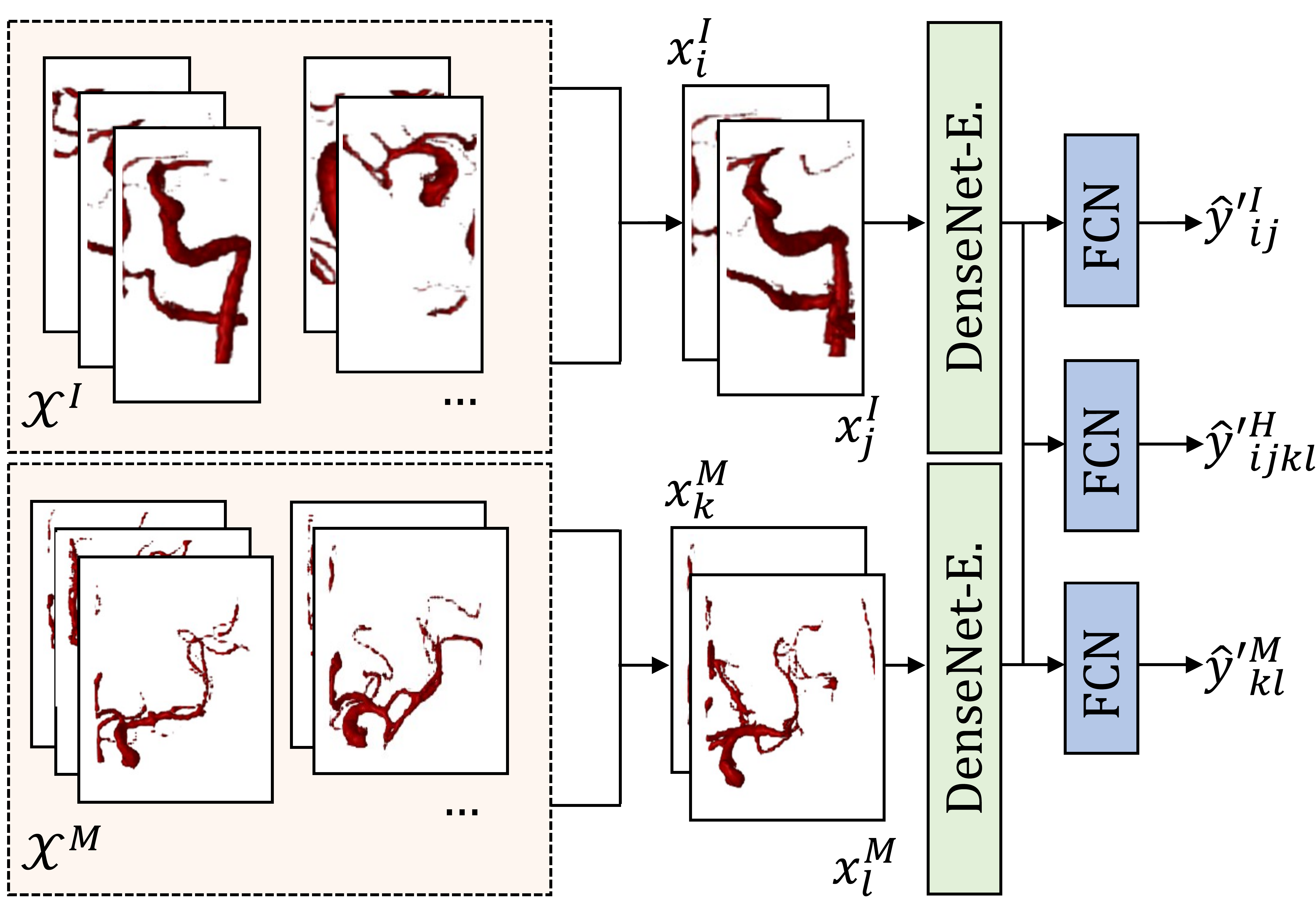}}
  \label{fig:networks}
    \caption{Schematics of the H-Stack and IM-Stack variant both predicting the presence of an LVO and in the positive case its side.}
\end{figure}

In the second variant, samples $s_{ijkl}$ are drawn from  $\mathcal{X}^{\text{I}}$ and $\mathcal{X}^{\text{M}}$ according to the scheme presented in Sec.~\ref{sec:recombination-im}. The ICA and MCA volumes are concatenated channel-wise with their side-counterpart as visualized in Fig.~\ref{fig:im-stack}. Each vessel region is encoded with a DenseNet ($\approx$\,3m parameters, growth rate of 16 and 32 initial feature maps) similarly as described in the H-Stack variant. The last latent space feature vectors (globally max pooled to a length of 64) of both encoders are forwarded to three individual fully connected heads (consisting of one dense layer and one batch-norm layer each) reducing the space to the required dimension of three. The global head receives a concatenation of both encodings. LVOs can affect either the ICA or MCA or both of the vessels. By analyzing the ICA and MCA regions separately, the network is able to focus on the characteristic LVO-patterns in one vessel, independent to any occurrences of LVOs in the respective other vessel. The ICA and MCA volumes are deformed as well (max.~displacement 20 voxels, 4 anchors, 10 repetitions).

\subsection{Experiments}
All models of Sec.~\ref{sec:architectures} are trained by Adam \cite{3214-kingma2014adam} (learning rate $10^{-5}$, batch size 6) using the BCE-Loss (PyTorch 1.6 \cite{3214-pytorch} and Python 3.8). We furthermore used the 3D-Densenet implementation by \cite{3214-densenet3d} and, besides the changes in Sec.~\ref{sec:architectures}, kept default values for all other details concerning the DenseNets used in this work. Early stopping (p=100 epochs) has been applied monitored on the validation AUC. To evaluate the effect of the recombination, we perform an ablation study combined with a 5-fold cross validation (3-1-1 split for training, validation and testing). Validation/Testing is done on original, not recombined, data. We applied the deformation (D) and recombination (R) on each method, including the baseline. Additionally, the data is mirrored (M) for the baseline.

\section{Results}
The AUCs for the ``Class'' categories, depicted in Tab.~\ref{tab:results}, are determined by evaluating the test data according to the probability of the LVO-pos.~class (sum of left and right LVO-pos.~class) against the LVO-neg.~class (no LVO). ``Side'' is measured by the accuracy of taking the argmax of the left or right class prediction on the LVO-pos.~cases. Potential false negatives are taken into account. Variances and 95\% confidence intervals can be found in the suppl.~material. It is evident that the recombination by itself consistently leads to a better performance for the classification of LVOs than deformation. Both augmentations combined complement each other boosting every method to its individual best performance. Except for the ``Whole Head'' model, which seems to not properly detect ICA LVOs while being accurate in predicting the side in all three classes. H-Stack detects ICA and MCA LVOs better than the baseline ``Whole Head'' except on the global label, where both methods perform equally well. In contrast, IM-Stack outperforms the other methods w.r.t.~the LVO classification by a significant margin, achieving an AUC of 0.91 globally, 0.96 for the ICA and 0.91 for the MCA, while for the side being superior or close to the best performer.

\begin{table}[tb]

\centering
\ra{1.3}
\caption{Quantitative evaluation of the 5-fold cross validation showing the ROC-AUCs of the respective class-wise prediction for the presence/absence of an LVO, and the accuracy for the affected side on LVO-pos.~cases. The abbreviation ``R'' stands for the proposed recombination method, ``D'' for the deformation and ``M'' for mirroring.}
\label{tab:results}

\begin{tabular}{@{}l>{\centering}m{10mm}<{\centering}m{10mm}>{\centering}m{10mm}<{\centering}m{10mm}>{\centering}m{10mm}<{\centering}m{10mm}@{}}
\toprule

                                      & \multicolumn{2}{c}{\hspace{-1em}Global} & \multicolumn{2}{c}{\hspace{-1em}ICA} & \multicolumn{2}{c}{\hspace{-1em}MCA} \\ \cline{2-7} 
Method & Class AUC & Side Acc. & Class AUC & Side Acc. & Class AUC & Side Acc.\\ \hline
Whole   Head  \cite{3214-thammlinger-bvm}       & 0.79             & 0.84    & 0.82            & 0.86  & 0.79            & 0.84  \\
Whole Head + D + M  \cite{3214-thammlinger-bvm} & 0.84             & 0.94    & 0.86            & 0.88  & 0.85            & 0.97  \\
Whole Head + R                         & 0.87             & 0.94    & 0.82            & 0.86  & 0.87            & 0.97  \\
Whole Head + R + D                     & 0.89             & 0.93    & 0.81            & \textbf{0.94}  & 0.89            & 0.98  \\ \hline
H Stack                                & 0.73             & 0.91    & 0.74            & 0.83  & 0.75            & 0.94  \\
H Stack + D                            & 0.82             & 0.86    & 0.86            & 0.86  & 0.85            & 0.93  \\
H Stack + R                            & 0.87             & 0.93    & 0.95            & 0.86  & 0.89            & 0.94  \\
H Stack + R + D                        & 0.89             & 0.92    & 0.95            & 0.88  & \textbf{0.91}   & 0.96  \\ \hline
IM Stack                               & 0.84             & 0.92    & 0.71            & 0.81  & 0.82            & 0.91  \\
IM Stack + D                           & 0.86             &\textbf{0.96}    & 0.93            & 0.88  & 0.86            & \textbf{0.99}  \\
IM Stack + R                           & 0.88             & 0.94    & 0.92            & 0.86  & 0.89            & 0.97  \\
IM Stack + R + D                       & \textbf{0.91}    & \textbf{0.96}    & \textbf{0.96}   & 0.92  & \textbf{0.91}   & 0.98  \\
\bottomrule
\end{tabular}
\end{table}

\section{Conclusion}

We present a novel and efficient augmentation technique for patient-level LVO classification based on the recombination of segmented subvolumes from multiple patients. While the data sets created in this manner may be of limited realism overall due to differences in patient anatomy, contrast bolus or image quality, their use in training consistently and significantly improved the performance of all tested models. It appears that repeatedly presenting the individual subvolumes to the models in new contexts strongly benefits generalization, even if the artificial data sets providing that context are less representative of real samples. We evaluated the proposed recombination as well as a state-of-the-art deformation-based augmentation in a baseline architecture and two models specifically designed exploiting the inherent symmetry of the brain to detect and coarsely localize (ICA/MCA) an anterior LVO. Best results were achieved when both types of augmentation were combined with our task-specific models.

\bibliographystyle{splncs04}
\bibliography{bibliography}

\begin{thebibliography}{10}
\providecommand{\url}[1]{\texttt{#1}}
\providecommand{\urlprefix}{URL }
\providecommand{\doi}[1]{https://doi.org/#1}

\bibitem{3214-amukotuwa2019fast}
Amukotuwa, S.A., Straka, M., Dehkharghani, S., Bammer, R.: Fast automatic
  detection of large vessel occlusions on ct angiography. Stroke
  \textbf{50}(12),  3431--3438 (2019)

\bibitem{3214-amukotuwa2019automated}
Amukotuwa, S.A., Straka, M., Smith, H., Chandra, R.V., Dehkharghani, S.,
  Fischbein, N.J., Bammer, R.: Automated detection of intracranial large vessel
  occlusions on computed tomography angiography: a single center experience.
  Stroke  \textbf{50}(10),  2790--2798 (2019)

\bibitem{caplan2007arterial}
Caplan, L.R.: Arterial occlusions: does size matter? Journal of Neurology,
  Neurosurgery \& Psychiatry  \textbf{78}(9),  916--916 (2007)

\bibitem{3214-chefd2002flows}
Chefd'Hotel, C., Hermosillo, G., Faugeras, O.: Flows of diffeomorphisms for
  multimodal image registration. In: Proceedings IEEE International Symposium
  on Biomedical Imaging. pp. 753--756 (2002). \doi{10.1109/ISBI.2002.1029367}

\bibitem{3214-densenet3d}
Hara, K., Kataoka, H., Satoh, Y.: Can spatiotemporal 3d cnns retrace the
  history of 2d cnns and imagenet? In: Proceedings of the IEEE Conference on
  Computer Vision and Pattern Recognition (CVPR). pp. 6546--6555 (2018)

\bibitem{3214-huang2017densely}
Huang, G., Liu, Z., Van Der~Maaten, L., Weinberger, K.Q.: Densely connected
  convolutional networks. In: IEEE Conference on Computer Vision and Pattern
  Recognition (CVPR). pp. 4700--4708 (2017). \doi{10.1109/CVPR.2017.243}

\bibitem{3214-kemmling2012decomposing}
Kemmling, A., Wersching, H., Berger, K., Knecht, S., Groden, C., N{\"o}lte, I.:
  Decomposing the hounsfield unit. Clinical neuroradiology  \textbf{22}(1),
  79--91 (2012)

\bibitem{3214-kingma2014adam}
Kingma, D.P., Ba, J.: Adam: A method for stochastic optimization (2015)

\bibitem{3214-luijten2021diagnostic}
Luijten, S.P., Wolff, L., Duvekot, M.H., van Doormaal, P.J., Moudrous, W.,
  Kerkhoff, H., a~Nijeholt, G.J.L., Bokkers, R.P., Lonneke, S., Hofmeijer, J.,
  et~al.: Diagnostic performance of an algorithm for automated large vessel
  occlusion detection on ct angiography. Journal of neurointerventional surgery
   (2021). \doi{10.1136/neurintsurg-2021-017842}

\bibitem{3214-pytorch}
Paszke, A., Gross, S., Massa, F., Lerer, A., Bradbury, J., Chanan, G., Killeen,
  T., Lin, Z., Gimelshein, N., Antiga, L., et~al.: Pytorch: An imperative
  style, high-performance deep learning library. Advances in neural information
  processing systems  \textbf{32} (2019)

\bibitem{3214-torchio}
P{\'e}rez-Garc{\'i}a, F., Sparks, R., Ourselin, S.: Torchio: a python library
  for efficient loading, preprocessing, augmentation and patch-based sampling
  of medical images in deep learning. Computer Methods and Programs in
  Biomedicine p. 106236 (2021).
  \doi{https://doi.org/10.1016/j.cmpb.2021.106236}

\bibitem{3214-stib2020detecting}
Stib, M.T., Vasquez, J., Dong, M.P., Kim, Y.H., Subzwari, S.S., Triedman, H.J.,
  Wang, A., Wang, H.L.C., Yao, A.D., Jayaraman, M., et~al.: Detecting large
  vessel occlusion at multiphase ct angiography by using a deep convolutional
  neural network. Radiology  \textbf{297}(3),  640--649 (2020)

\bibitem{3214-thammlinger-vcbm}
Thamm, F., Jürgens, M., Ditt, H., Maier, A.: {VirtualDSA++: Automated
  Segmentation, Vessel Labeling, Occlusion Detection and Graph Search on
  CT-Angiography Data}. In: Kozlíková, B., Krone, M., Smit, N., Nieselt, K.,
  Raidou, R.G. (eds.) Eurographics Workshop on Visual Computing for Biology and
  Medicine. The Eurographics Association (2020). \doi{10.2312/vcbm.20201181}

\bibitem{3214-thammlinger-bvm}
Thamm, F., Taubmann, O., Jürgens, M., Ditt, H., Maier, A.: Detection of large
  vessel occlusions using deep learning by deforming vessel tree segmentations.
  In: Bildverarbeitung in der Medizin. Springer, Heidelberg (2022)

\end{thebibliography}

\end{document}